\documentclass{ws-procs9x6}

\begin{document}

\title{Hadron attenuation by (pre)hadronic FSI at HERMES}

\author{T.~Falter$^{1}$, K.~Gallmeister$^{2}$ and U.~Mosel$^{2}$}

\address{
$^{1}$Physics Department, Brookhaven National Laboratory, USA\\ 
$^{2}$Institut f\"ur Theoretische Physik, 
Universit\"at Gie\ss en, Germany}  

\maketitle

\abstracts{
We investigate hadron production in deep inelastic lepton-nucleus scattering in the kinematic regime of the HERMES experiment. Our calculations are carried out in the framework of a BUU transport model which contains the Lund event generators PYTHIA and FRITIOF for the simulation of high-energy elementary interactions. For the first time we consistently use the complete four-dimensional information of the Lund string break up vertices as input for our transport theoretical studies of (pre)hadronic final state interactions. We compare our results with experimental HERMES data on charged hadron attenuation.}

\section{Introduction}
Hadron production in deep inelastic lepton-nucleus scattering (DIS) provides an ideal tool to investigate the space-time evolution of hadron formation\cite{Kop}. The nuclear target can be viewed as a kind of 'micro-detector' that is located directly behind the virtual photon-nucleon interaction vertex. It allows us to study the interactions of the reaction products with the surrounding nuclear environment on a length scale that is set by the size of the target nucleus. By comparison with hadron production on a deuterium target one can draw conclusions about the space-time picture of hadronization. The latter information is crucial for the interpretation of jet quenching in ultra-relativistic heavy ion-collisions at RHIC as a possible signature for the creation of a deconfined quark-gluon plasma phase\cite{Cassing}. 

In the recent past the HERMES collaboration at DESY has started an extensive experimental investigation of hadron attenuation in DIS on various gas targets\cite{HERMES} and a similar experiment is currently performed at Jefferson Lab\cite{JLab}. Furthermore, hadron attenuation in nuclear DIS will be subject to investigation after the 12 GeV upgrade at Jefferson Lab and at a possible future electron-ion collider\cite{eRHIC}. In Refs.~\refcite{partonic} the observed attenuation of high energy hadrons at HERMES has been interpreted as being due to a partonic energy loss of the quark that was struck by the virtual photon. The colored quark undergoes rescattering inside the nucleus giving rise to induced gluon radiation. However, the authors of Refs.~\refcite{prehadron} and ourselves\cite{Falter,Fal04} achieve a very good description of the experimental data by assuming that the struck quark forms a color neutral prehadron early after the virtual photon-nucleon interaction. This prehadron then scatters off the surrounding nucleons on its way out of the nucleus. 

In Refs.~\refcite{Falter}, \refcite{Fal04} and references therein we have developed a method to incorporate coherence length effects in a semi-classical Boltzmann-Uehling-Uhlenbeck (BUU) transport model that uses the Lund string models PYTHIA\cite{Pythia} and FRITIOF\cite{Fritiof} for the simulation of the elementary DIS process and the (pre)hadronic final state interactions (FSI) respectively. This allows for a complete probabilistic coupled-channel description of high-energy photo- and electroproduction off complex nuclei. Our simulation works on an event-by-event basis and can be directly compared to experiment accounting for all sorts of kinematic cuts and detector acceptances. Our theoretical results are in perfect agreement with the experimental findings at HERMES.

\section{Model}
In our model we split the lepton-nucleus interaction into two parts: In step 1) the exchanged virtual photon interacts with a bound nucleon inside the nucleus and produces a final state that in step 2) is propagated within the transport model. We take into account nuclear effects such as binding energies, Fermi motion and Pauli blocking. We also account for shadowing of the resolved photon components using the method developed in Refs.~\refcite{Shadowing}. 

\begin{figure}[t]
	\begin{center}
    \includegraphics[width=9cm,angle=-90]{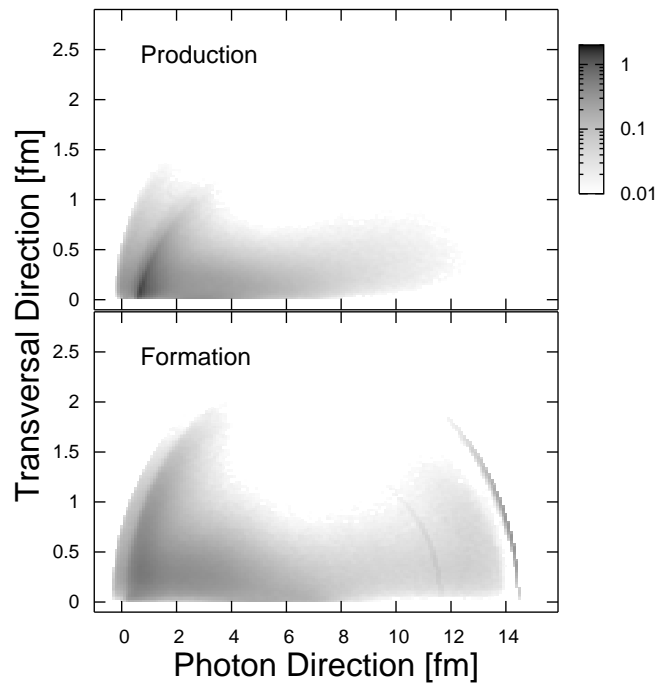}
	\end{center}
	\caption{Production (top) and formation (bottom) points for a typical Hermes event: (photon energy $\nu=14$ GeV,  virtuality $Q^2=2.5$ GeV$^2$). The target nucleon is located at the origin, the virtual photon is coming in from the left. \label{fig:points}}
\end{figure}
The final state of the initial photon-nucleon reaction is determined by PYTHIA, i.e.~the direct and resolved photon-nucleon interactions lead to the excitation of one or more hadronic strings which fragment according to the Lund fragmentation scheme\cite{Lund}. Each time a quark-antiquark pair is produced in the fragmentation process a string splits into two color neutral fragments. If such a color neutral fragment has the mass of a hadron we call it a prehadron. After the formation time the hadronic wave-function has build up and the prehadron has turned into a hadron with the same energy and momentum. In Ref.~\refcite{Formation} we have developed a method to extract the complete space-time information of each string fragmentation in PYTHIA. Although our results should not be overstressed since one applies a semi-classical picture to a quantum mechanical problem, our approach makes it possible to assign the four-dimensional production and formation point to each single hadron in each single scattering event. Figure~\ref{fig:points} shows the spatial distribution of the prehadron production and hadron formation points for a typical HERMES event. Obviously, a large fraction of these points fall within a distance behind the struck nucleon that is comparable to nuclear radii.

Starting from their production point the prehadrons propagate through the nucleus and interact with the surrounding nuclear matter. In this work we set their interaction cross section to the corresponding hadronic value. The propagation is described within our semi-classical BUU transport model\cite{Fal04,Eff99} which allows for a probabilistic coupled-channel description of the FSI. Each time a particle interacts with a bound nucleon it might either scatter elastically or produce new particles that are also propagated. At high energies the final state of an inelastic collision is determined by the event generator FRITIOF which is also based on the Lund fragmentation scheme. Finally, we end up with a complete lepton-nucleus event that can be corrected for experimental cuts and detector acceptance.

\section{Results}
The observable of interest is the so-called multiplicity ratio
\begin{equation}
\label{eq:multiplicity-ratio}
R_M^h(z_h,\nu)=\frac{\frac{N_h(z_h,\nu)}
{N_e(\nu)}\big|_A}{\frac{N_h(z_h,\nu)}{N_e(\nu)}\big|_D}
,
\end{equation}
where $N_h$ is the yield of semi-inclusive hadrons in a given $(z_h,\nu)$-bin and $N_e$ the yield of inclusive deep inelastic scattering leptons in the same $\nu$-bin. The quantity $z_h=E_h/\nu$ denotes the energy fraction of the hadron. For the deuterium target, i.e.~the nominator of Eq.~(\ref{eq:multiplicity-ratio}), we simply use the isospin averaged results of a proton and a neutron target. Thus in the case of deuterium we neglect the FSI of the produced hadrons and also the effect of shadowing and Fermi motion. 

\begin{figure}[t]
	\begin{center}
    \includegraphics[width=12cm]{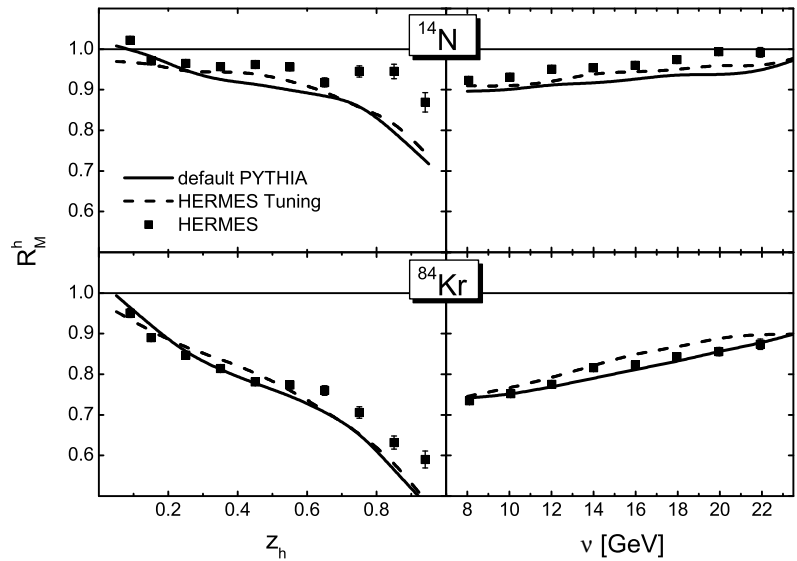}
	\end{center}
	\vspace{-0.75cm}
	\caption{Calculated multiplicity ratio of charged hadrons for $^{14}$N and $^{84}$Kr nuclei using the four-dimensional production points from the PYTHIA model and setting the prehadronic cross section to the hadronic value. The solid line shows the result using the default PYTHIA parameters. For the calculation represented by the dashed line the fitted parameter set of Ref.~\protect\refcite{Patty} has been used. The data are taken from Ref.~\protect\refcite{HERMES}. \label{fig:NKr}}
\end{figure}

Figure~\ref{fig:NKr} shows the charged hadron multiplicity ratio for a 27.6 GeV positron beam incident on a $^{14}$N and $^{84}$Kr target. In our calculations we assume the same kinematic cuts on the scattered lepton and produced hadrons as in the HERMES experiment. In addition we take into account the geometrical acceptance of the HERMES detector. The solid curve shows the result using the default PYTHIA parameter set. In the calculation represented by the dashed curve we used the PYTHIA parameters that have been fitted to hydrogen data by the HERMES collaboration\cite{Patty}. Both calculations yield comparable results that are in good agreement with the experimental data.

\section{Conclusions}
Our results demonstrate that one needs large prehadronic cross sections to describe the experimentally observed hadron attenuation in nuclear DIS at HERMES when using the prehadron production points from PYTHIA. In this work we have simply set the prehadronic cross sections to the corresponding hadronic values. So far we have neglected all FSI of the string and string fragments prior to the production of the prehadrons and assumed that string fragmentation in nuclei does not differ from that in vacuum. Both effects might increase the hadron attenuation\cite{prehadron}. In future work one might also think of incorporating induced gluon radiation prior to the production of the color neutral prehadrons. However, one should prevent double counting when considering both gluon bremsstrahlung and the FSI of the hadronic string.

\section*{Acknowledgments}
Work supported by BMBF. T.F. is supported by the Alexander von Humboldt Foundation (Feodor Lynen Research Fellowship).


\begin{thebibliography}{0}

\bibitem{Kop} 
B.~Kopeliovich, J.~Nemchik, and E.~Predazzi, 
in {\it Proceedings of the workshop on Future Physics at HERA}, 
edited by G.~Ingelman, A.~De Roeck, R.~Klanner, DESY, 1995/96, vol 2, p. 1038, nucl-th/9607036.

\bibitem{Cassing}
K.~Gallmeister, C.~Greiner, Z.~Xu,
Phys.~Rev.~C67 (2003) 044905;
W.~Cassing, K.~Gallmeister, and C.~Greiner, 
Nucl.~Phys.~A {\bf 735}, 277 (2004);
J.~Phys.~G {\bf 30}, S801 (2004).

\bibitem{HERMES}
A.~Airapetian et al.~[HERMES Collaboration],
Eur.~Phys.~J.~C {\bf 20}, 479 (2001);
Phys.~Lett.~B {\bf 577}, 37 (2003).

\bibitem{JLab}
W.~K.~Brooks, 
Fizika B {\bf 13}, 321 (2004).

\bibitem{eRHIC}
A.~Deshpande, R.~Milner, R.~Venugopalan and W.~Vogelsang,
hep-ph/0506148.

\bibitem{partonic} 
E.~Wang, X.-N.~Wang, 
Phys.~Rev.~Lett. {\bf 89}, 162301 (2002); 
F.~Arleo, 
Eur.~Phys.~J. C {\bf 30}, 213 (2003).

\bibitem{prehadron}
A.~Accardi, V.~Muccifora and H.-J.~Pirner, 
Nucl.~Phys.~A {\bf 720}, 131 (2003);
B.~Z.~Kopeliovich, J.~Nemchik, E.~Predazzi and A.~Hayashigaki,
Nucl.~Phys.~A {\bf 740}, 211 (2004);
A.~Accardi, D.~Grunewald, V.~Muccifora and H.~J.~Pirner,
hep-ph/0502072.

\bibitem{Falter}
T.~Falter, W.~Cassing, K.~Gallmeister and U.~Mosel, 
Phys.~Lett.~B {\bf 594}, 61 (2004). 

\bibitem{Fal04}
T.~Falter, W.~Cassing, K.~Gallmeister and U.~Mosel,
Phys.~Rev.~C {\bf 70}, 054609 (2004).

\bibitem{Pythia}
T.~Sj\"ostrand, P.~Eden, C.~Friberg, L.~L\"onnblad, G.~Miu, S.~Mrenna, E.~Norrbin, 
Comp.~Phys.~Commun.~135 (2001) 238;
T.~Sj\"ostrand, L.~L\"onnblad and S.~Mrenna, 
LU TP 01-21 [hep-ph/0108264].

\bibitem{Fritiof}
Hong Pi,
Comput.~Phys.~Commun.~{\bf 71}, 173 (1992);
B.~Andersson, G.~Gustafson, and Hong Pi,
Z.~Phys.~C {\bf 57}, 485 (1993).

\bibitem{Shadowing}
M.~Effenberger and U.~Mosel,
Phys.~Rev.~C {\bf 62}, 014605 (2000);
T.~Falter and U.~Mosel,
Phys.~Rev.~C {\bf 66}, 024608 (2002);
T.~Falter, K.~Gallmeister, and U.~Mosel,
Phys.~Rev.~C {\bf 67}, 054606 (2003).

\bibitem{Lund}
B.~Andersson, G.~Gustafson, G.~Ingelman, T.~Sj\"ostrand,
Phys.~Rept.~97 (1983) 31; 
B.~Anderson, in:
{\it The Lund Model}, Cambridge University Press (1998).

\bibitem{Formation}
K.~Gallmeister and T.~Falter, Phys.~Lett.~B (in press),
nucl-th/0502015.

\bibitem{Eff99}
M.~Effenberger, E.~L.~Bratkovskaya, and U.~Mosel,
Phys.~Rev.~C {\bf 60}, 44614 (1999).



\bibitem{Patty}
Patricia Liebing, Ph.D. Thesis, 2004, University of Hamburg.

\end{thebibliography}
\end{document}